\documentclass[12pt,preprint]{aastex}
\shorttitle{Photometry and Spectroscopy of KS UMa}

\shortauthors{Zhao et al.}
\begin{document}
\title{Photometry and Spectroscopy of KS Ursae Majoris during Superoutburst}
\author{Yinghe Zhao, Zongyun Li, Xiaoan Wu, Qiuhe Peng}
\affil{Department of Astronomy, Nanjing University, Nanjing
210093, China} \email{yhzhao, zyli@nju.edu.cn} \and
\author{Zhousheng Shang, Zili Li}
\affil{Yunnan Observatory, Kunming 650011, China}

\begin{abstract}

We report photometric and spectroscopic observations of the SU
UMa-type dwarf novae, KS Ursae Majoris, during its 2003 February superoutburst.
Modulations with a period of $0.07017\pm0.00021$ day, which is
3.3\% larger than the orbital period, have been found during the
superoutburst and may be positive superhumps. A maximum
trough-to-peak amplitude of around 0.3 magnitude is determined for
this superhump.

The spectra show broad, absorption-line profiles. The lines
display blue and red troughs which alternate in depth. The radial
velocity curve of the absorption wings of H$\beta$ has an
amplitude of $40\pm11$ km s$^{-1}$ and a phase offset of
$0.12\pm0.03$. The $\gamma$ velocity of the binary is $3\pm9$ km
s$^{-1}$ and varies on an order of 50 km s$^{-1}$ from day to day.
From another clear evidence for a precessing eccentric disk, we
obtain a solution to an eccentric outer disk consistent with
theoretical works, which demonstrates the validity of the relation
between superhumps and tidal effects. The inner part of the disk
is also eccentric as evidenced by asymmetric and symmetric wings
in the lines. Therefore, the whole disk is eccentric and the
variation of $\gamma$ velocity and the evolutionary asymmetric
line profiles could be criterions for an precessing eccentric
accretion disk.

\end{abstract}


\section{Introduction}

SU Ursae Majoris stars, which comprise a subgroup of dwarf novae
(DN), were defined with their distinctive superoutbursts and
accompanying superhump phenomena (for a review, see Warner 1985,
1995). Superhumps are large amplitude luminosity variations with
period, $P_{sh}$, slightly displaced from the orbital period
$P_{orb}$. The most common type occur that $P_{sh}$ is a few
percent longer than $P_{orb}$, thought to arise from the
interaction of the second star orbit with a slowly progradely
precessing non-axisymmetric accretion disk, as explained as the
result of the ``thermal-tidal instability" (Osaki 1989, hereafter
TTI). The eccentricity of the disk arises because a 3:1 resonance
occurs between the second star orbit and motion of matter in the
outer disk. This can only occur in systems that with sufficiently
low mass ratio ($q=M_2/M_1$) that the 3:1 resonance radius is
within the tidal radius at which the disk is truncated by tidal
forces (Paczy\'{n}ski 1977).

KS UMa was discovered in the Second Byurakan Sky Survey (Markarian
\& Stepanian 1983) and identified as a cataclysmic variable (CV)
star by Balayan (1997).
It identified as an SU UMa-type CV by Hazen \& Garnavich
(1999). Its luminosity varies between $V\sim 12.5$ mag and $V\sim
17.1$ mag. The unpublished materials by Vanmunster (1998) and
Nogami (1998) suggested that KS UMa had a superhump period of
0.06875 day, and Patterson et al. (2003) obtained a $P_{orb}$ of
0.06796 day from radial curve of H$\alpha$ line measured in
quiescence.

So far there is no spectrocopic and detailedly photometric study
on KS UMa when it is going through superoutburst. In this paper,
we report the results of V-band photometry and spectroscopy of KS
UMa during the superoutburst occurred 2003 February. Our
observations revealed a precise superhump period, 101 min, and
evidences (e.g. asymmetric line profiles and variable $\gamma$
velocities) for a precessing eccentric accretion disk. This paper
is organized as follows: in Section 2 we describe our observations
both for photometry and spectroscopy, and present the results and
analysis in Section 3. In Section 4 we discuss our results, and
present a summary in the last Section.

\section{Observations}

We took photometric observations of KS UMa for 60 hrs over 6 nights
(from 2003 February 21 to February 26, UT), using a TEK1024 CCD
camera attached to the Cassegrain focus of the 1.0 m reflector at
Yunnan Observatory. Exposure time was set long enough to assure
good signal-to-noise ratio.  A total 1270 useful object frames
were obtained through the $V$ filter. After bias subtraction and
flat field correction, we removed the sky background. Differential
magnitudes of KS UMa were obtained by using 2 secondary
photometric standards, stars 2 and 5 on the finding chart,
KSUMA-CCDF.PS, which was downloaded from the web page of
AAVSO\footnote{ http://www.aavso.org/}, as the comparison star and
check star, respectively. The rms error of magnitude was less than
0.02 mag.

The spectroscopic observations were conducted with the
Optomechanics Research, Inc., Cassegrain spectrograph attached to
the 2.16 m telescope with a TEK1024 CCD camera at Xinglong Station
of the National Astronomical Observatory. The technique of
observation and data processing is similar to that in Wu et al.
(2001, named Paper I hereafter). A 600 groove mm$^{-1}$ grating
was used, and the slit width was set to 2$''$.5. Dome flats were
taken at the beginning and end of each night. Exposure time for
the star ranged from 900 to 1200, depending on weather conditions.
Total observational time was 14.4 hr, 8.8 times of the orbital
period. Twenty and twenty-three star spectra were collected on
February 25 and 26 (Beijing time), respectively. The lamp spectra
recorded before and after every two successive star exposures were
used to interpolate the coefficients of the wavelength scales. We
derived a spectral resolution of 8 \AA \ from FWHM measurement of
the lamp spectra. The rms error of identified lines was less than
0.15 \AA \ using a fourth-order Legendre polynomial to fit the
lines, corresponding to 9 km s$^{-1}$ near H$\beta$. A detailed
observation journal is summarized in Table 1.

\section{Results and Analysis}

\subsection{Photometric Period and Mean Magnitude}
Figure 1(a) shows the entire light curves of our observations. The
sumperhump with a peak-to-trough amplitude around 0.3 mag can be
seen clearly in the light curves, and their periodic nature is
well established in all six days. Figure 1(c) and 1(d) show its
periodgram and window spectrum (Scargle 1982) for the whole
observations, respectively. The powerful signal is at a frequency
of 14.24816 cycles$\cdot$day$^{-1}$, corresponding to a period of
$P_{sh}=0.07017\pm0.00021$ day. Figure 1(b) shows the data, which
have been subtracted the mean brightness of every day and are
folded with the period of 0.07017 day. This is an obvious pattern
of rapid rise and slow decline, as is usually found in the common
superhumps of dwarf novae. The superposing solid line in Figure
1(b) is the best-fit sinusoid. This sinusoid gives an ephemeris,
\begin{displaymath}
 T_{0}=HJD2452692.11267+0.07017E
\end{displaymath}
where $T_{0}$ represents the time of hump maximum, $E$ is the
cycle number. We also computed periodgram and window spectrum for
individual data series. The results of period determination and
full amplitudes of magnitude variation(${\Delta}$${m}$, derived
from fitting sinusoidal to data in Figure 1(a)) are also listed in
Table 2. It is not unreasonable to find the evolution of the
superhump period with date (Warner 1995) in KS UMa. Furthermore,
it has the similar behavior to OY Car, whose period increased for
the first few days of superoutburst (Krzeminski \& Vogt 1985) and
then decreased in the late stage of superoutburst (Schoembs 1986).

\subsection{Radial Velocity}
In Figure 2(A) we show the sum of 20 individual spectra obtained
on the first night and Figure 2(B) shows the sum of 23 individual
spectra on the second night. These two spectra have been
normalized to the continuum. When combined, no radial velocity
shift was applied. The most prominent features are the Balmer
absorption lines, with obvious emission component seen in the core
of H$\beta$. There is also He~I $\lambda$4471 line presenting in
the spectra.

The RV curve of absorption wings has two privileges over that of
absorption cores for our spectra. First, the spectra present asymmetric ``W"
patterns, so it is hard to obtain the true wavelengths of the cores. Second,
since absorption wings ($\Delta V$=1000-2000 km\ s$^{-1}$) come
from the inner disk, the RV of wing component represents the
motion of the white dwarf much better than the core component.
We used the double-Gaussian convolution method (Shafter et al.
1988) to measure the RV of the wings; the $\sigma$ of each
Gaussian is set as 500 km s$^{-1}$. Figure 3 shows the diagnostic
diagram (Thorstensen et al. 1991) of H$\beta$, which is not
blended with other lines. The orbital phase was computed according
to the ephemeris given by Patterson et al. (2003), $T_0=HJD
2,451,332.6724(11)+0.06796E$, where $T_0$ is the time of the
$\gamma$ crossover from negative to positive velocities and $E$ is
a cycle number.

Figure 3 suggests that $\sigma_{K}/K$ decreases and increases
gradually, before and after a separation of 1950 km s$^{-1}$,
respectively. So we adopt $K$ as $40\pm11$ km s$^{-1}$ and
$\gamma$ as $3\pm9$ km s$^{-1}$, at the separation of 1950 km
s$^{-1}$. These results are well consistent with $47\pm5$ km
s$^{-1}$ and $8\pm3$ km s$^{-1}$ measured in quiescence (Patterson
et al. 2003). The radial curve fitted with a least-squares
sinusoidal is shown in Figure 4.

\subsection{Mass and Inclination}

Patterson (2001) found a relation between the period excess,
$\epsilon$, and the mass ratio, $q$ (=$M_{2}/M_{1}$) by fitting
some data of ($\epsilon, q$). The relation is written as
\begin{displaymath}
\epsilon=0.216(\pm0.018)q
\end{displaymath}
where $\epsilon=(P_{sh}-P_{orb})/P_{orb}$. Using the orbital
period $P_{orb}=0.06796(10)$ day (Patterson et al. 2003), and the
superhump period $P_{sh}$=0.07017(21) day, we get
$\epsilon$=0.0325(35). This gives $q=0.15\pm0.02$.

Assuming the secondary is a main-sequence star, $M_{2}\approx
M_{\odot}(R_{2}/R_{\odot})^{5/4}$, according to an empirical
mass-radius relation (Kippenhahn \& Weigert 1990). Because the
secondary fills its Roche lobe, $R_{2}\approx R_{L}(2)=aQ(q)$,
where $Q(q)=0.49q^{2/3}/[0.6q^{2/3}+~\ln(1+~q^{1/3})]$ (Warner
1995). Considering $P_{orb}=0.06796(10)$ day (Patterson et al.
2003), we obtain $a=0.7037(1+~1/q)^{1/3}M_{2}^{1/3}R_{\odot}$.
Thus
\begin{equation}
M_1=M_2/q,\ M_2=[0.7037(1+1/q)^{1/3}Q(q)]^{15/7}
\end{equation}

The mass of the white dwarf should be less than 1.44$M_{\odot}$;
it only requires that $q>0.06$ according to equation (1). The mass
ratio of 0.15$\pm0.02$ derived in this paper is consistent with
this requirement. If all assumptions made here are correct, we can
obtain that $M_{2}=0.09\pm0.01M_{\odot}$ and
$M_{1}=0.60\pm0.07M_{\odot}$. The mass function $f(M)=(M_{2}\sin
i)^3/(M_1+M_2)^2]=K_1^3P_{orb}/(2\pi G)=0.00073(10)\ M_\odot$
gives $i=51^\circ\pm12^\circ$. However, the masses given here are
rather uncertain because we have made some unproved assumptions.

\subsection{An Eccentric Disk}
\subsubsection{Features of the Absorption Line Profiles}
Figure 5 shows the profile evolution of H$\beta$ through orbital
phase during the two days. Qualitatively, the blue and red troughs
seemed to alternate moving up and down, in a ``see-saw" pattern.
The red trough is \emph{always} deeper on February 25, while the
blue one is \emph{always} deeper on February 26.

In Figure 6 we show the mean profiles of H$\beta$ of the two days,
which have normalized to the continuum intensity. The abscissas
are relative wavelengths to the central wavelength, which
represents the $\gamma$ velocity. The dotted lines are their
images with wavelength inverted. The wings (Blue Wing of the Blue
absorption component (BWB) and Red Wing of the Red absorption
component (RWR)) are absolutely asymmetric within $\Delta \lambda$
$\sim$ 15-35 \AA\ and the BWB is \emph{always} shallower than the
RWR on February 25. However,the situation became more complex on
February 26, as shown in the top right panel in Figure 6. The RWR
within $\Delta \lambda$ $\sim$ 15-25 \AA\ is shallower than the
BWB but they became almost symmetric within $\Delta \lambda$
$\sim$ 25-35 \AA. In addition, we examined every phase-binned
spectrum and found the same features mentioned above for all of
the spectra. The difference among the phase-binned spectra
obtained on February 26 is that the range of asymmetry (or
symmetry) is different according to the orbital phase.

Table 3 lists the measured wavelengths of the troughs of H$\beta$
by fitting the lower four or five pixels with a parabola, their
difference $(R-B)$ and mean $(R+B)/2$. Note that the wavelengths
of the two troughs both shorter on February 25 than on February
26, while the difference is small (less than 1 angstroms). EWs of
H$\beta$ and the wavelengths of the ``emission" peaks are also
measured (see Table 3). Similar to the troughs, the wavelength of
``emission" peak is shorter on February 25 than on February 26. We
also noticed that the redshift of the ``emission" peak (2.8 \AA)
is much larger than that of the mean wavelength of the troughs
(0.6 \AA). The mean wavelength of all troughs is well consistent
with that of ``emission" peaks, which is 4861.5 \AA.

The similar phenomena has been found in the spectra of AM Canum
Venaticorum (Patterson et al. 1993, named Paper II hereafter). The
authors proposed a simple model, which is on the basis of an
apsidal precessing eccentric disk, to explain this phenomena.
Paper I proposed a detailed method, which is based on a similar
model, to deal with asymmetry in emission lines of IY UMa during
superoutburst. We processed our data with the similar method
proposed by Paper I, the small difference is that Paper I used
$\Delta V_{peak}$ while we used $\Delta V_{trough}$ to calculate
the relation between $\Theta$, the longitude of the periastron for
the first day (the angle between the line of sight and the major
axis of the elliptical disk), and the eccentricity $e$. The mean
velocity of the emission peaks or absorption troughs for each day
is $V=-C e \sin \Theta$ (Paper I), where $C=\sin i
\sqrt{GM_1/a(1-e^2)}=constant$; $i$ and $a$ are the inclination
and half of the major axis of the accretion disk, respectively.
Therefore, $V=-C e \sin \Theta$ and $V=-C e \sin (\Theta+2.92)$
for Feb 25 and Feb 26, respectively. The increase of 2.92 rad of
the longitude of the periastron for the second day is due to the
precession of the disk. Using the value of ``Redshif of troughs"
listed in Table 3, we find,
\begin{equation}
\cos(\Theta-\pi+1.46)=0.041/e
\end{equation}
where we have adopted $M_1$ and $i$ as the values calculated in
Section 3.2, the precessing period as 2.16 day, and $a$ is assumed
to be 0.52 times as the length of the half of the major axis of
the binary orbit, according to equation (2.61) in Warner (1995).
So $e$ must be equal to or larger than 0.041. We derive two
solutions for every $e$ of 0.041-0.21, each falling in one field
of (A): $\Theta=1.68\sim 3.06$ rad, and (B): $\Theta=1.68\sim
0.31$ rad. However, only solution B is correct (see~ $\S$ 3.4.2).

The ``mean of all troughs" should be equal to $4861.3+[\gamma -
Ce\sin(\Theta+2.92)/2-Ce\sin(\Theta)/2]/62$, where 62 km s$^{-1}$
is the radial velocity corresponding to 1 \AA\ reddening near
H$\beta$ and 4861.3 \AA\ is the stationary wavelength of H$\beta$.
Thus, considering equation (2) and solution B, we can obtain
$\gamma=17\pm6$ km s$^{-1}$. This value is consistent with $8\pm3$
measured in quiescence (Patterson et al. 2003) and in
superoutburst (see $\S$ 3.2).

\subsubsection{A Simple Model of Asymmetric Absorption Lines}
To study whether an eccentric disk can account for all of the
observed features, we computed the profiles of H$\beta$ from the
model which were used in Paper I and Paper II. We found that this
model worked well within the outer part of the disk (i.e., cores
of the absorption line) but poorly within the inner part of the
disk (i.e., wings of the absorption line). The emissivity was
taken as $r^{-\beta}$ both in Paper I and Paper II, for the
emission line and absorption line, respectively. In paper I,
$\beta$ is 1.2 for the inner disk and 2.1 for the outer disk, and
in Paper II $\beta$ is 1-2. Since these values are working poorly
for our observational data within the inner disk, we tried some
other values and found that $\beta = -1.2$ for the inner disk
works well. For outer disk, we took $\beta$ as 2.1, the same as
used in Paper I. The ratio of the outer disk to the inner disk was
set as 4 and 5-10, in Paper I and Paper II, respectively. In our
calculation, we adopted 10.

We have made a crude estimate for the emissivity as following: the
emissivity for H$\beta$ should be proportional to the number
density of the neutral hydrogen which is at the second excited
state. Using the Saha's formula and after some steps of
calculation, we have
\begin{equation}
 \frac{{N_{0,2} }}{{N_H }} = \frac{{N_e g_{0,2} h^3 e^{(\chi _0  - \varepsilon _{0,2} )/kT} }}
 {{N_e u_0 h^3 e^{\chi _0 /kT}  + 2u_1 (2\pi m_e kT)^{3/2} }}
  = \frac{{4 N_e e^{3.95/T_4 } }}{{N_e e^{15.6/T_4 }  + 2.4 \times 10^{21} T_4^{3/2} }} \\
  \end{equation}
where $N_{0,2}$, $N_H$ and $N_e$ are the number density of the
neutral hydrogen which is at the second excited state, the total
number density of hydrogen and the electronic number density,
respectively; $g_{0,2}$ the degree of degeneracy; $u_0$ and $u_1$
are the partition functions for neutral and ionized hydrogens,
respectively; and $T_4 = T/10^4$ K, $T$ is the temperature. At the
last step of calculation, we have substituted 13.6 eV for the
ionization potential of neutral hydrogen, $\chi _0$, and 10.2 eV
for the excited potential of neutral hydrogen at 2nd excited
state, $\varepsilon _{0,2}$.  For an accretion disk, $T$ can be
written as (Frank et al. 2002):
\begin{equation}
T = \left\{ {\frac{{3GM\dot{M} }}{{8\pi r^3 \sigma }}\left[ {1 -
\left( {\frac{{R_* }}{r}} \right)^{1/2} } \right]} \right\}^{1/4}
\end{equation}
where $R_*$, $M$ and $\dot{M}$ are the radius, mass and accretion
rate of the white dwarf, respectively; $r$ is the radius of the
disk; $\sigma$ the Stefan-Boltzmann constant and $G$ the
gravitational constant.

Using equation (3) and (4) and adopting the typical values for
white dwarfs, we did numerical calculations for $N_{0,2}$. We
found that the results can be fitted with a power law function
with the form of $r^{-\beta}$, where $\beta$ is negative for the
inner region but positive for the outer region. However, the
absolute value of $\beta$ is sensitive to $N_e$.

The lower two panels in Figure 6 are the sample-simulated H$\beta$
profiles, which have normalized to the continuum spectrum
intensity. We have set $e=0.17$, $\Theta=0.35$ to satisfy equation
(2). The sample-simulated profiles resemble the observed ones in
the following aspects. (1) The core of the blue absorption
component (CB) is \emph{obviously} shallower than the core of the
red absorption component (CR) on February 25, while the CR is
\emph{obviously} shallower on February 26. (2) The BWB is clearly
shallower on February 25, while part of the RWR is shallower on
February 26. (3) Part of the RWR and BWB become symmetric on
February 26. (4) The ``emission core" is blue shift on February 25
but red shift on February 26. Note that the simulated line
profiles are much narrower than the observed ones. This is because
various intrinsic broadening mechanisms (Marsh 1987, Horne 1995).

With the emissivity assumed above, we computed the line profiles
for different $e$ and $\Theta$ that satisfied equation (2). Figure
7 displays simulated dependence of the ratio of the
\emph{shallower} core of the absorption component to the
\emph{deeper} one on $\Theta$ during two days. The property that
the ratio is smaller on February 26 than on February 25 constrains
that $\Theta$ falls within 1.68 to 0.31. Therefore only solution B
(see section 3.4.1) is correct.

\section{Discussion}
\subsection{The superhump period}
The present observations have thus confirmed KS UMa as being a SU
UMa-type dwarf novae with well-established superhump morphology
and period. Using the empirical $\epsilon -~ P_{orb}$ relation
(Thorstensen et al. 1996),
\[\epsilon=-0.0344+(0.0382\ hr^{-1})\times P_{orb}\]
where $\epsilon=(P_{sh}-P_{orb})/P_{orb}$. Substituting the
orbital period with 0.06796 day (Patterson et al. 2003) and
combining these two equations, we can expect that the superhump
period is to be $\sim$0.06986 day, which is 2.8\% bigger than the
orbital period and is consistent with our result of 0.07017(21)
day.

\subsection{The evidence for an eccentric disk}
The TTI model was proposed to explain the bimodal outbursts of SU
UMa stars (Osaki 1989). Many numerical simulations (Hirose \&
Osaki 1990; Whitehurst 1994; Kunze et al. 1997; Murray 1998; Truss
et al. 2001) showed that thermal-tidal instability model was very
successful on explaining behaviors of SU UMa stars. The TTI model
requires that the accretion disk is eccentric and precessing.
However, it is difficult to study directly from observation
whether the accretion disk is eccentric and precessing or not.

There are some  groups who have made efforts in this research
region (Vogt 1982; Honey et al. 1988; Paper II; Paper I). Vogt
(1982) and Honey et al. (1988) have found that the $\gamma$
velocities of Z Cha varied during its superoutburst. Vogt (1982)
proposed a model in which he considered the behavior of a
precessing, elliptical ring surrounding a circular accretion disk.
This gives the variation of the $\gamma$ velocity on a
night-to-night basis as a result of variations in the projected
motion of the ring material against that of the inner (circular)
disk. Honey et al. (1988) interpreted their observational result
with new non-axisymmetric disk simulations as arising in an
eccentric, precessing disk which is tidally distorted by the
secondary. Paper II found the profiles change in helium lines of
AM Canum Venaticorum on a time scale of tens hours. The authors of
Paper II suggested that the changes of absorption line profiles
could be caused by a eccentric precessing disk. They also
suggested this phenomena would be seen in the emission line
profiles.

Paper I presented spectroscopic observations on IY UMa and found
evolution of asymmetric emission line profiles of H$\alpha$ and
H$\beta$. They showed that a slowly precessing eccentric accretion
disk could produce such asymmetric emission line profiles.

Our spectroscopic study on KS UMa presents another object for
evidences of a precessing eccentric accretion disk. As described
in Section 3.4, we reproduced the evolutional asymmetric profiles
for the absorption lines on the basis of the method proposed by
Paper I and the similar assumptions used in Paper I and Paper II,
except the index of the emissivity for the inner disk. This may be
caused by the following reasons: 1), Paper I is for the emission
line; 2), Paper II do not apply their simulated profiles to
analyse the features of symmetry/asymmetry, hence they may not
find it out whether the value of $\beta$ they used is reasonable
or not.

Moreover, we found that the $\gamma$ velocity varied from day to
day, as shown in Figure 8. The difference between these two
$\gamma$ velocities is on the order of $\sim$50 km s$^{-1}$
(corresponding to $\sim$ 0.8 \AA\ at H$\beta$). The systematic
error would not result in this value since we have checked the sky
emission 5577 \AA\ lines of these two days and no difference
larger than 0.1 \AA\ was found. We also found that the spectra of
Z Cha, as shown in Figure 6 in Vogt (1982) and in Figure 5 in
Honey et al. (1988) presented the asymmetric line profiles.
Therefore, it is not unreasonable to believe that the evolutionary
asymmetric line profiles and the variation of $\gamma$ velocity
could be criterions for an eccentric precessing accretion disk.

\section{Summary}

(1) The photometric data shows that the superhump period of KS UMa
is 0.07017$\pm$0.00021 day, which is 3.3\% lager than the orbital
period.

(2) The value of $K$ of the H$\beta$ absorption wings is 41$\pm$11
km s$^{-1}$, with a phase offset of 0.012$\pm$0.03. The $\gamma$
velocity of the binary is 3$\pm$9 km s$^{-1}$. The amplitude of
$K$ and $\gamma$ are consistent with those measured in quiescence.
But the $\gamma$ velocity varied on an order of $\sim$50 km
s$^{-1}$ from day to day.

(3) The mass ratio of the binary is 0.15$\pm$0.02. And we get that
$M_1$ and $M_2$ are 0.60$\pm$0.07 $M_\odot$ and 0.09$\pm$0.01
$M_\odot$, respectively. Therefore we obtain that the inclination
of the system is $51^\circ \pm 12^\circ$.

(4) The asymmetry of the absorption line profiles through out the
orbital phase and the redshift of the troughs of H$\beta$ clearly
shows that the accretion disk is eccentric and precessing. With
detailed analysis on the base of a coherent model, we get that the
the eccentricity of the disk must be large than 0.041 and we
present a simulative result with $e=0.17$.

\acknowledgments The authors are very grateful to the anonymous
referee for his/her careful reading of the manuscript and
thoughtful comments. We would like to thank the Optical Astronomy
Laboratory, Chinese Academy of Science and Jianyan Wei of the
National Astronomical Observatory and Peisheng Cheng of Yunnan
Observatory for scheduling the observations. This work is
supported by grant 10173005 and 10221001 from the National Natural
Science Foundation of the People's Republic of China.

\begin{deluxetable}{ccccc}
\tablewidth{0pt}
\tablecaption{Journal of photometry and
spectroscopy}
\tablehead
{\colhead{Date (UT)} &\colhead{HJD Start} &\colhead{HJD End} &\colhead{Exposure} &\colhead{ }\\
\colhead{(Year 2003)}&\colhead{(2,452,000+)} &
\colhead{(2,452,000+)}  & \colhead{(s)} &\colhead{Exposures}\\
\hline\noalign{\smallskip} \multicolumn{5}{c}{Photometry}}
 \tablecolumns{5}
 \startdata
Feb 21 ........& 692.0401 & 692.4252 & 200 & 141\\
Feb 22 ........& 693.0519 & 693.4336 & 100 & 226\\
Feb 23 ........& 694.0074 & 694.4444 & 100 & 246\\
Feb 24 ........& 695.0095 & 695.4022 & 100 & 218\\
Feb 25 ........& 696.0141 & 696.4108 & 100 & 210\\
Feb 26 ........& 697.0102 & 697.4125 & 100 & 229\\
\hline\noalign{\smallskip}
\multicolumn{5}{c}{Spectroscopy}\\
\hline\noalign{\smallskip}
Feb 25 ........& 696.1238 & 696.3948 & 900,1200 & 20\\
Feb 26 ........& 697.0671 & 697.3965 & 900,1200 & 23\\
\enddata
\end{deluxetable}
\begin{deluxetable}{ccccc}
\tablewidth{0pt} \tablecaption{Nightly means of brightness,
variable rates, superhumps periods and amplitudes}
\tablehead{
 Date     & Mean brightness          & Variable rate       & Period    & $\Delta$m\\
 (UT)          & mag                   &  mag d$^{-1}$        & day       & mag
}
\tablecolumns{5} \startdata
 Feb 21       & 12.72         & -0.075$\pm$0.051 &0.07020$\pm$0.00128  & 0.30-0.20\\
 Feb 22       & 12.83         & -0.184$\pm$0.053 &0.07045$\pm$0.00130  & 0.25-0.20\\
 Feb 23       & 12.99         &0.206$\pm$0.026 &0.07086$\pm$0.00115  & 0.25-0.15\\
 Feb 24       & 13.14         &0.263$\pm$0.035 &0.07073$\pm$0.00127  & 0.22-0.15\\
 Feb 25       & 13.28         &0.068$\pm$0.033 &0.07024$\pm$0.00124  & 0.25-0.15\\
 Feb 26       & 13.35         &0.014$\pm$0.027 &0.06890$\pm$0.00118  & 0.25-0.15\\
 Feb 21-26    & 13.07         &0.132$\pm$0.001 &0.07017$\pm$0.00021  & 0.30-0.15\\
 \enddata
 \end{deluxetable}

\begin{center}
\begin{deluxetable}{lcccccc}
\tablewidth{0pt}
\tabletypesize{\footnotesize}
\tablecaption{Parameters of H$\beta$$\ \lambda$4861} \tablehead
 {\colhead{Date (UT)} &\colhead{EW} &\colhead{Red trough} &\colhead{Blue trough} &\colhead{$(R+B)/2$}
 &\colhead{$R-B$} &\colhead{``Emission" peak}\\
\colhead{(2003)} &\colhead{(\AA)}&\colhead{(\AA)} &\colhead{(\AA)}
&\colhead{(\AA)} &\colhead{(km s$^{-1}$)} &\colhead{(\AA)}}
\tablecolumns{8} \startdata
Feb 25 ..........         & 5.1        &4872.3       &4850.1      & 4861.2    & 1370   &4860.1 \\

Feb 26 ..........        &5.0         &4873.1       &4850.5      & 4861.8    & 1395   &4862.9 \\

\hline\noalign{\bigskip}
...     &...  &... & Mean of all  &Redshift of troughs     &Mean of all &Redshift of peaks   \\
...     &...   &...&troughs (\AA)  &(km s$^{-1}$)          &peaks (\AA)  &(km s$^{-1}$)      \\
\hline\noalign{\bigskip}
Feb 25 \& 26    &... &... &4861.5          & 37            &4861.5 &173 \\
\enddata
\end{deluxetable}
\end{center}

\clearpage

\begin{figure}
\figurenum{1} \epsscale{1.0} \plotone{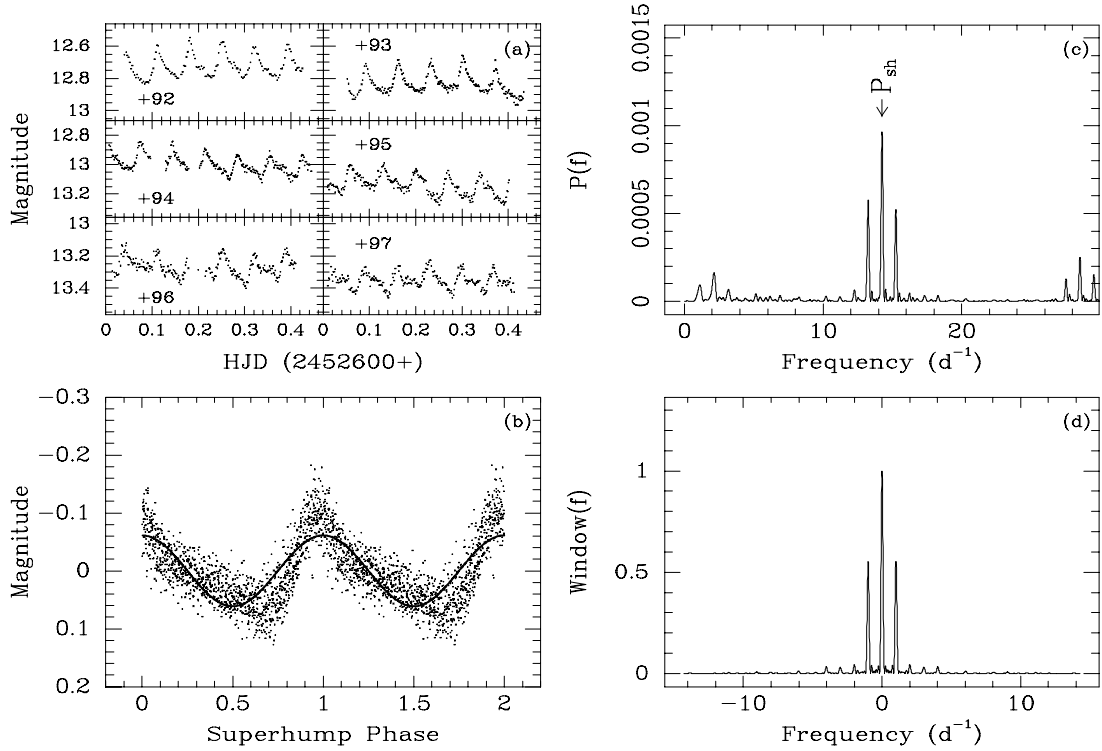}
\caption{Photometric data and its periodgram. (\emph{a}) All data.
The numbers in the subwindows are what should be added to the
horizontal axis. (\emph{b}) Folded data with a period of 0.07017
day. The mean magnitude and the trend were removed. (\emph{c}) The
power spectrum. (\emph{d}) The window spectrum.}
\end{figure}

\begin{figure}
\figurenum{2} \epsscale{1.0} \plotone{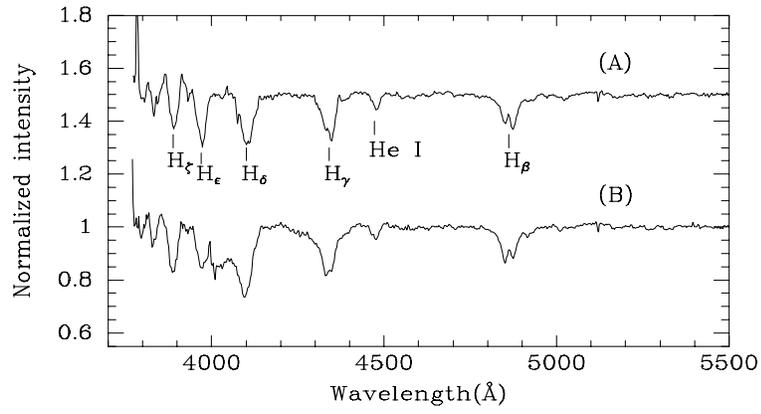} \caption{Normalized
average spectra of KS UMa during superoutburst. (A): observed on
February 25; (B): observed on February 26. The spectrum show the
prominent Balmer absorption lines with emission components and
weaker HeI lines.}
\end{figure}

\begin{figure}
\figurenum{3} \epsscale{1.0} \plotone{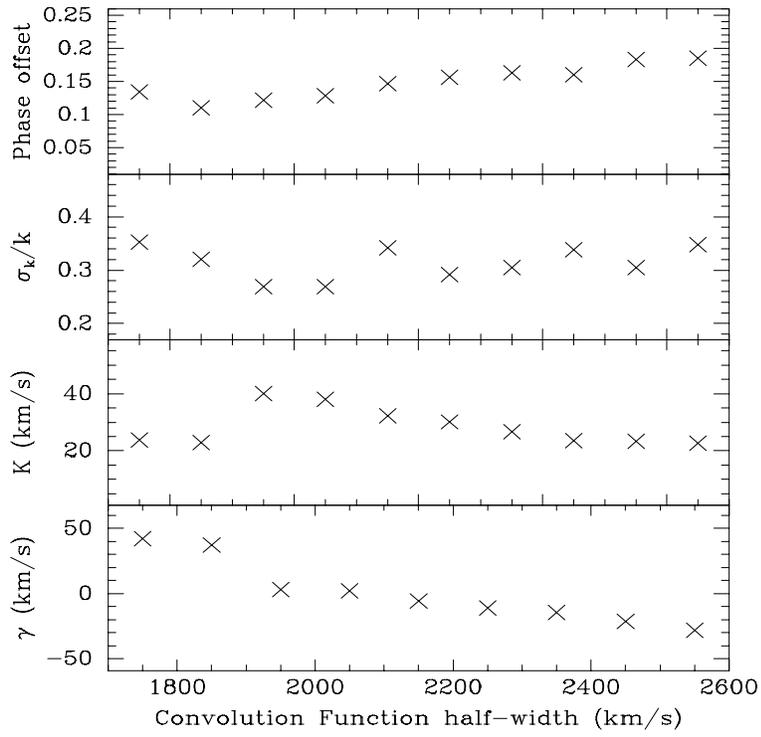}
\caption{Diagnostic diagram based on the H$\beta$ observations.
The data are obtained using the double-Gaussian convolution
method.}
\end{figure}

\begin{figure}
\figurenum{4} \epsscale{1.0} \plotone{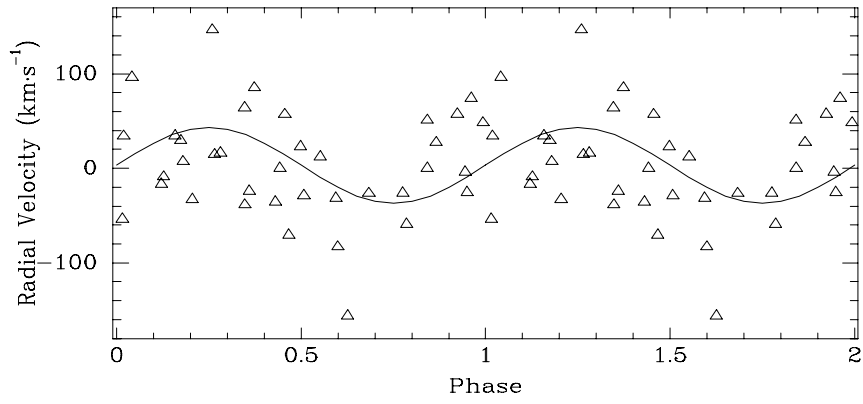}
\caption{Least-squares sinusoidal fitted for the radial velocities
of the wings of H$\beta$ of February 25 and 26.}
\end{figure}

\begin{figure}
\figurenum{5} \epsscale{1.0} \plotone{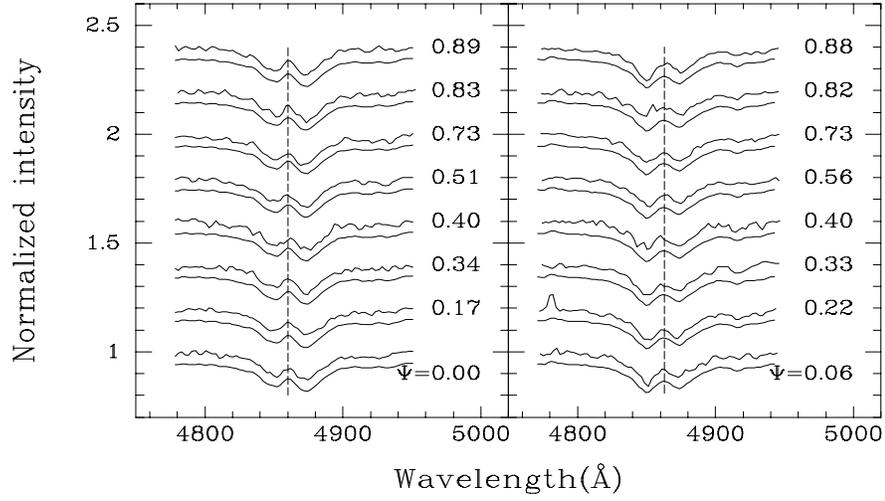} \caption{Evolution
of the normalized H$\beta$ profile through the orbital phase.
\emph{Left:} February 25. \emph{Right:} February 26. The dashed
lines show the center of the emission cores of the average
spectra, which are 4860.1 \AA\ and 4862.9 \AA, respectively. The
lower line in every pair lines is the intraday average spectrum.}
\end{figure}

\begin{figure}
\figurenum{6} \epsscale{1.0} \plotone{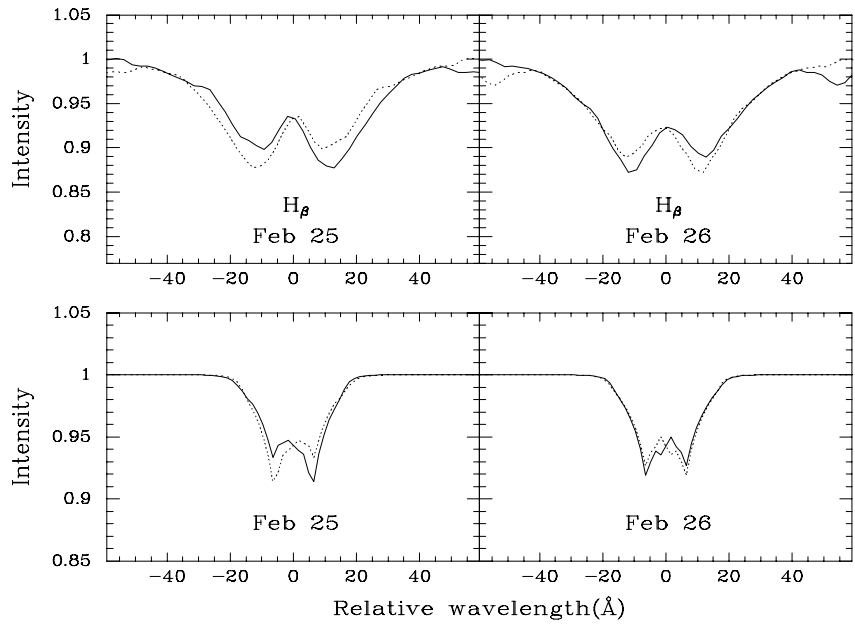} \caption{The upper
two panels show the asymmetry and symmetry of H$\beta$ on February
25 and 26. The asymmetry is exactly opposite for the two days. The
lower two panels show two sample-simulated H$\beta$ profiles,
which are produced according to the model in the text. They
reproduced all the asymmetric and symmetric features of the
observed profiles.}
\end{figure}

\begin{figure}
\figurenum{7} \epsscale{1.0} \plotone{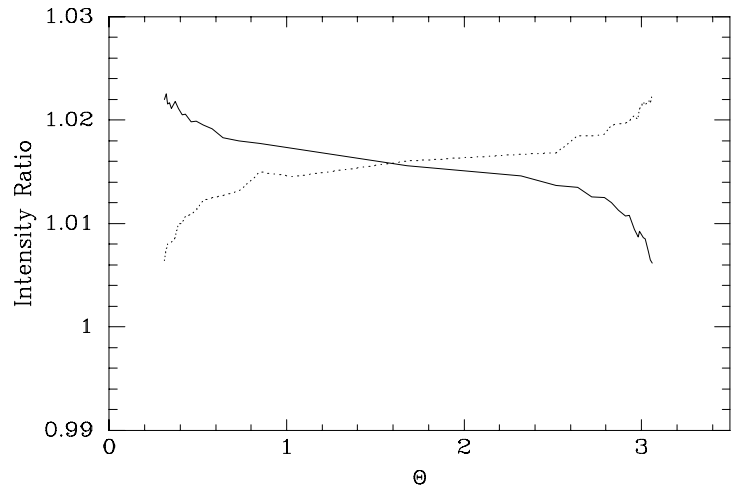} \caption{Simulated
dependence of the ratio of the \emph{shallower} core of the
absorption component to the \emph{deeper} one on $\Theta$. The
solid and dotted lines are for February 25 and 26, respectively.
From the property that the ratio is factually larger on February
25, we know that only solution B is correct.}
\end{figure}

\begin{figure}
\figurenum{8} \epsscale{1.0} \plotone{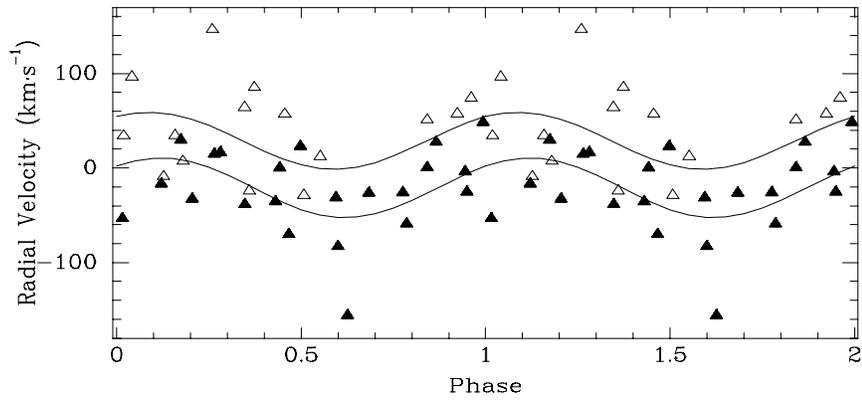}
\caption{Least-squares sinusoidal fitted separately for the radial
velocities of the wings of H$\beta$ of February 25 (open
triangles) and 26 (solid triangles). It shows clearly that the
$\gamma$ velocity of these two days varied with a amplitude of
$\sim$50 km s$^{-1}$.}
\end{figure}

\end{document}